\documentclass[12pt]{article}

\usepackage{amssymb} 
\usepackage{amsmath} 
\usepackage{graphicx}
\usepackage{hyperref}
\usepackage{color,soul}

\usepackage[margin=0.5in]{geometry}
\usepackage{times}

\begin{document}

{\bf  \large
Using local plasticity rules to train recurrent neural networks
}\\
{
Owen Marschall, Kyunghyun Cho, and Cristina Savin 
}\\


{\bf Summary:} To learn useful dynamics on long time scales, neurons must use plasticity rules that account for long-term, circuit-wide effects of synaptic changes. In other words, neural circuits must solve a  \emph{credit assignment problem} to appropriately assign responsibility for global network behavior to individual circuit components. Furthermore, biological constraints demand that plasticity rules are spatially and temporally \emph{local}; that is, synaptic changes can depend only on variables accessible to the pre- and postsynaptic neurons. While artificial intelligence offers a computational solution for credit assignment, namely \emph{backpropagation through time} (BPTT),
this solution is wildly biologically implausible. It requires both nonlocal computations and unlimited memory capacity, as any synaptic change is a complicated function of the entire history of network activity. Similar nonlocality issues plague other approaches such as FORCE \cite{sussillo2009generating}. Overall, we are still missing a model for learning in recurrent circuits that both works computationally and uses only local updates. Leveraging recent advances in machine learning on approximating gradients for BPTT, we derive biologically plausible plasticity rules that enable recurrent networks to accurately learn long-term dependencies in sequential data. The solution takes the form of neurons with segregated voltage compartments, with several synaptic sub-populations that have different functional properties. The network operates in distinct phases during which each synaptic sub-population is updated by its own local plasticity rule. Our results provide new insights into the potential roles of segregated dendritic compartments, branch-specific inhibition, and global circuit phases in learning.\\
\begin{figure}[h]
\includegraphics[width=7.5in]{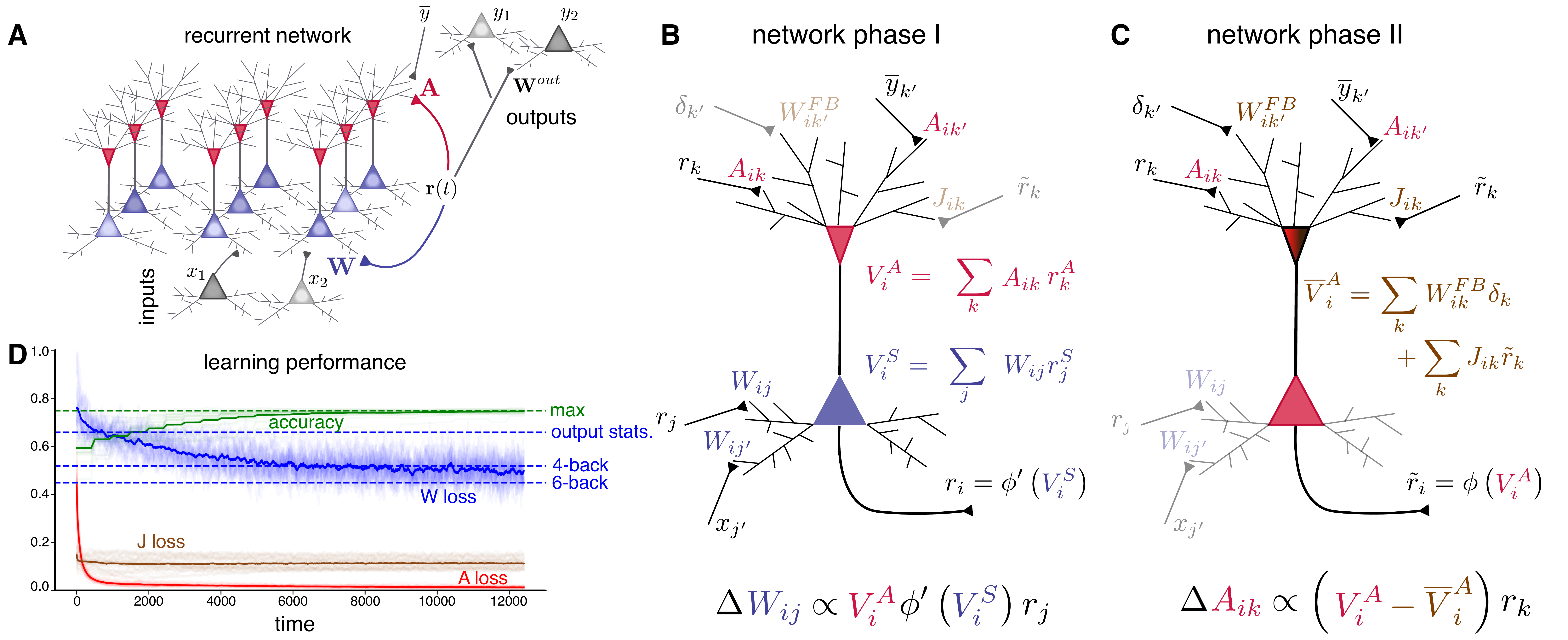}
\caption{{\bf (A)} The network architecture consists of just one hidden layer, receiving inputs from an input population ${\bf x}$ and projecting to output units ${\bf y}$. {\bf (B)}: Basal phase of network activity, during which $\phi = \tanh$ determines the firing rate $r_i$ from the somatic voltage $V^S_i$. The vectors ${\bf r}^S$ and ${\bf r}^A$ are concatenations of ${\bf r}$ with the inputs ${\bf x}$ and labels ${\bf \overline{y}}$, respectively, with a constant $1$ to provide a bias term. {\bf (C)}: Distal phase of network activity, during which neural firing $\tilde{r}_i$ is determined by $\phi\left(V_i^A\right)$. {\bf (D)}: Temporally filtered losses for each set of synapses. Test accuracy (green) was computed every 500 time steps. Light traces are individual trials, while dark traces are trial-averages.}
\end{figure}

We chose the (4, 6)-back task to quantify learning \cite{pitis2016recurrent} because it has low-dimensional inputs and outputs, multiple time scales of relevant information, and clear bounds for performance that correspond to learning particular input-output dependencies. In more detail, the network has to map an i.i.d. temporal sequence of Bernoulli inputs with $p_x = 0.5$ to a Bernoulli output, whose probability depends on the inputs with some lag that can be adjusted to tune the task difficulty, here lags of 4 and 6 time steps. In particular, the baseline output probability $p_y = 0.5$ is increased (decreased) by 0.5 (0.25) when the input from 4 (6) time steps back is equal to 1.\\
 
 Our model consists of an input layer, a recurrent network, and an output layer (Fig.1A). We define plasticity rules with the aim of minimizing a loss function that quantifies task performance as the cross entropy between the network outputs and the target distribution. Plasticity rules can be derived by performing stochastic gradient descent on this loss function, but calculation of the exact gradient requires computations that are nonlocal over space and time. Instead, we exploit a novel machine learning technique, known as ``synthetic gradients," to approximate the gradient using local computations \cite{jaderberg2017decoupled}. Biologically, this approximation manifests as a network of neurons with multiple compartments that are innervated by functionally distinct sets of synapses ${\bf W}$, ${\bf A}$ and ${\bf J}$, one somatic and two distal. The ${\bf W}$ synapses are used for solving the actual task, processing inputs and running the primary network dynamics, whereas the ${\bf A}$ and ${\bf J}$ synapses are used for learning. All of these synapses are plastic.\\
 
 First, the plasticity rule for a synapse $W_{ij}$ has a very simple form, depending only on presynaptic activity and the somatic and distal postsynaptic voltages $V_i^S$ and $V_i^A$ (Fig.1B). Biologically, this corresponds to distal modulation of synaptic plasticity \cite{dudman2007role}.
 Second, the plasticity rule for a synapse $A_{ij}$ depends on presynaptic activity and postsynaptic distal voltage, gated by the inputs from the ${\bf J}$ synapses and a top-down error signal $\boldsymbol{\delta}$ passed through fixed feedback synapses ${\bf W^{FB}}$ (Fig.1C).
 Lastly, the plasticity rule for $J_{ij}$ follows perceptron-like learning $\Delta J_{ij} \propto r_i^{(t)} \left(r_j^{(t+1)} - \sum_k J_{jk} r_k^{(t)}\right)$ to implement one-step prediction of recurrent dynamics. ${\bf J}$ is meant to approximate the Jacobian of the recurrent dynamics $\partial {\bf r}^{(t+1)} /\partial {\bf r}^{(t)}$. Since all the required signals cannot be simultaneously represented at the level of voltages, we require two distinct phases for the network dynamics: a ``somatic" phase for the updates of ${\bf W}$ and ${\bf J}$, and a ``distal" phase for the updates of ${\bf A}$. Biologically, this could be mediated by targeted inhibition that dynamically gates out unwanted inputs \cite{somogyi2014temporal}.\\
 
 Each plasticity rule minimizes an implicit loss function w.r.t. its synaptic sub-population: ${\bf W}$ updates to improve task performance, ${\bf A}$ to improve the approximation of credit assignment, and ${\bf J}$ to approximate the Jacobian. Fig.1D shows the evolution of the corresponding losses over learning. As expected, the losses decrease and stabilize. Interestingly, the saturation happens fastest for ${\bf J}$, driving learning in ${\bf A}$, which in turn drives learning in ${\bf W}$, predicting possible differences in timescales of plasticity at distal vs. basal synapses. What does the network learn? Performance-wise, the network produces the correct output $75\%$ of the time, which is the theoretical bound given inherent randomness in the task. The 3 blue dashed lines represent cross-entropy bounds for ``internalizing" the different dependencies between the inputs and outputs. The upper-most dashed line represents learning of the marginal output statistics, i.e. that $62.5\%$ of outputs are 1, while the second and third dashed lines represent learning of 4- and 6-time-step lags, respectively. On average over many random seeds, our model reliably learns the 4-time-step lag and is sufficiently close to the next bound to indicate some knowledge of the 6-back component. Failing to learn the second dependency is not entirely surprising, because optimal performance w.r.t. cross entropy requires a perfect calibration of confidence at each time step. Moreover, vanilla recurrent networks are known to struggle with long-term dependencies even with full BPTT \cite{pascanu2013difficulty}. The fact that our local approximation of a much more complicated algorithm can learn long-term dependencies at all is exciting.\\
 
 In summary, we have designed a network model that can learn long-term dependencies using biologically plausible, local learning rules. The required biological features for calculating credit assignment include multi-compartment neurons, distinct phases for circuit dynamics, and spatial clustering of synapses with similar function. While functional roles of different compartments and their distinct plasticity properties have received experimental attention, there is relatively little theoretical work on their computational significance \cite{guerguiev2017towards}. Our work is an important step in this direction.
 
\bibliographystyle{ieeetr}
{\footnotesize
\bibliography{ref}}
 
\end{document}